\documentclass[prl]{revtex4}
\usepackage{amsmath}
\usepackage{graphicx}
\newcommand{\avg}[1]{\left\langle{#1}\right\rangle}
\newcommand{\ba}{\begin{eqnarray}}
\newcommand{\ea}{\end{eqnarray}}
\newcommand{\bi}{\begin{itemize}}
\newcommand{\ei}{\end{itemize}}
\newcommand{\ii}{\item}
\newcommand{\la}{\label}
\newcommand{\f}{\frac}
\newcommand{\nn}{\nonumber\\}
\def\abar{\bar \alpha_S(L)}
\def\yb{\tilde y}
\def\ka{\kappa}

\def\g{\gamma}
\def\lam{\lambda}
\def\b{\beta}
\def\al{\alpha}
\def\creation{\lambda}
\def\bx{\bar X}
\def\r{\rho}
\begin{document}
\title{Extended Geometric Scaling from Generalized Traveling Waves}

\author{R. Peschanski}
\email{robi.peschanski@cea.fr}
\affiliation{Institut de Physique Th{\'e}orique (IPhT), CEA/Saclay,
  91191 Gif-sur-Yvette cedex, France\footnote{
URA 2306, unit\'e de recherche associ\'ee au CNRS.}}

\begin{abstract}
We define a  mapping of  the QCD Balitsky-Kovchegov equation
 in the  diffusive approximation   with noise and a generalized coupling allowing a common
 treatment of the fixed and running QCD couplings. 
 It corresponds to the  extension of the stochastic Fisher 
and Kolmogorov-Petrovsky-Piscounov equation  to the  radial wave 
propagation in a medium with  negative-gradient absorption 
responsible for anomalous diffusion,
non-integer dimension and damped noise fluctuations. We obtain its analytic 
traveling wave solutions with a  new scaling curve and for running coupling 
a new scaling variable allowing to extend the range and validity of the 
geometric-scaling QCD prediction beyond the previously known domain.
\end{abstract}

\maketitle


\section{1. Introduction}
\la{Intro}

Geometric scaling is originally \cite{geomsc} an empirical scaling law satisfied 
by the data on deep-inelastic totally inclusive cross-sections.
It may be translated as a property of the  scattering amplitude 
\ba
T(L,Y)= T\left[\tau(L,Y)\right]
\la{geom}
\ea
where the scaling variable $\tau(L,Y)=L-vY$  in the original formulation. The 
usual kinematic variables are $Y,$ the total rapidity and 
$L=\log(Q^2/\Lambda^2)$ with $Q$ the photon virtuality and $\Lambda$ the Quantum 
Chromodynamics (QCD) reference scale at one-loop level.

On the theoretical ground, there exists known encouraging features of QCD in the  saturation regime leading to approximate 
geometric scaling, but the situation is not yet settled. The good news were, for 
instance, that  in the large-$N_c$ limit and in the 
mean-field approximation, 
we are led to consider the Balitsky-Kovchegov (BK) equation \cite{bk} which 
possesses  the nice property \cite{mp} to be mapped, in the diffusive  
approximation, onto the Fisher-Kolmogorov-Petrovsky-Piscounov (F-KPP) equation 
\cite{fkpp,bramson} . This equation
which has been widely studied in statistical physics  is known to admit 
traveling waves as asymptotic solutions which translate into geometric scaling in 
QCD \cite{geomsc}. In fact one may enlarge that result to a whole class of 
equations, including the one considering the full QCD kernel \cite{bfkl} beyond 
the restriction to the first two derivatives. One may introduce the notion of 
``universality class'' since the asymptotic properties of the solutions, 
including the approximate geometric scaling, do not depend either on initial conditions 
or on  a precise form of the nonlinear part of the equation.

However, many questions remain unsolved. On a theoretical ground, there are 
quite a few problems to ensure analytical scaling properties, even in the simpler 
case of BK equation with running coupling, see $e.g.$ the discussion of 
Ref.\cite{Beuf}. Even more puzzling is the discovery  that introducing a small 
cut-off \cite{Brunet} to the FKPP or BK equation or adding a small white noise 
\cite{fluct}, i.e. for a stochastic (s)FKPP or (s)BK   equation, which after all 
reflects a physical constraint relying on the finiteness and randomness of the 
number of gluons, predicts a strong violation of geometric scaling not seen on 
data. On the other hand, the same stochasticity applied to the running coupling
case 
seems from indirect evidence either to preserve geometric scaling 
\cite{Dumi} or to delay its violation to much higher energy \cite{Beuf1}.

So, apart from the phenomenological problems, which we will discuss in a further 
publication,  the aim of the present  paper  is on the general theoretical issue of geometric scaling from the BK nonlinear evolution equation, in particular  for the running coupling case. 
Translating the problem in more mathematical terms, it seems that the 
``universality class'' of the BK equation with runnning coupling has not yet  
been found or at least clarified. It is our goal to propose a solution to this 
problem, and explore its consequences which happen to give rise to an extended 
scaling property, as will become 
clear soon. In fact we will define the initial mathematical problem  as  the 
diffusive approximation of the (s)BK equation with  generalized running 
coupling, extrapolating from the running or non-running case. 

One  starts 
by considering the following  one-parameter family
\begin{eqnarray}
L^n\ \partial_Y N(L,Y)  =  \left\{A_2 \partial_L^2 + A_1 \partial_L + A_0 
\right\}\ N(L,Y) -  N^2(L,Y)+  \sqrt{\ka N(L,Y)} \nu(L,Y) \ ,
                  \label{math}
\end{eqnarray}
where the case $n\!=\!1\ (resp.\ n\!=\!0)$ corresponds to the logarithmically 
running (resp. non-running) coupling  in the  (s)BK equation within the diffusive 
approximation ($1/{L}^n$ has been divided  
on both sides of \eqref{math}). Universality properties, independent from initial conditions and the form of the nonlinear damping,  will provide the suitable identification of the universality class. In equation 
\eqref{math}
\begin{equation}
\chi(-\partial_L)\sim A_2 \partial_L^2+ A_1 \partial_L +  A_0 
        \label{kernel0}
\end{equation}
represents for our purposes the expansion up to two derivatives of the linear 
kernel, either for the FKPP equation (where $\chi(\g) = \g^2+1$) or for the 
diffusive approximation of the Balitsky, Fadin, Kuraev and Lipatov (BFKL) kernel 
\cite{bfkl} relevant for the BK equation. The strength of the noise term is 
specified by
 the phenomenological fudge factor $\ka,$ and  $\nu(L,Y)$ is the standard 
white noise satisfying $\avg{\nu}=0$ and $
\avg{\nu(L,Y)\nu(L',Y')} =  \delta(Y-Y')\delta(L-L')\ .$

Investigating the properties of the general nonlinear evolution equation 
\eqref{math}, the  paper is organized as follows. In section 2, we show 
that Eq.\eqref{math}  can be exactly mapped onto the  extension of the 
stochastic Fisher and Kolmogorov-Petrovsky-Piscounov equation for the  radial wave propagation in a medium with absorptive negative gradient, non-integer 
dimension and damped noise fluctuations when $n>0$. In section 3, we derive 
new scaling solutions for arbitrary $n$ 
verifying the general recipe \eqref{geom},  defining for every value of $n$ a  scaling variable $\tau(L,Y)$ 
and a new scaling curve $T(\tau(L,Y))$ valid in the forward front of the 
traveling wave in a region (in $L$)  beyond but neighbouring the 
previously considered \cite{Brunet,mp} domain. By compatibility with those
previous universal results, in section 4 we establish the universality forms, 
indexed by $n,$ of the traveling wave solutions and focus on the physically 
interesting cases $n\!=\!0,1.$ We check recovering known results for 
$n\!=\!0,1$ and find a  geometric scaling parameterization extended in a larger kinematic domain  and
for the running case  a new scaling variable. Summary and outlook can be found in section 5.

\section{2. Mapping}
\la{Mapp}

The  initial idea of our approach is to introduce a change of variables
aiming at restoring the ``standard'' constant diffusion term, which determines 
the mathematical order of the (linear part of the) partial differential equation 
\eqref{math}. In some sense one is guided by restoring the classical diffusion equation (or also ``heat equation'')
${\partial}_T  = \partial_X^2 +\cdots,$ where $T$ is ``time''and $X$ is ``space''. One then writes the Ansatz
\ba
L=X^{\b}\ ;\quad \partial_L = \f 1\b\ {X^{1-\b}}\ \partial_X\ ,
\la{varia}
\ea
with $\b$ to be determined for our purpose.

After a straightforward transformation, Eq.\eqref{math} takes the 
form
\ba
X^{n\b}\ \f{\partial N(X,Y)}{\partial Y}  = \left\{\f{A_2}{\b^2} 
X^{2-2\b}\partial_X^2+ \left(A_2\f{1\!-\!\b}{\b^2} X^{1-2\b}+\f{A_1}{\b}  
X^{1-\b}\right) \partial_X + A_0  \right\} N(X,Y)-
\nn 
-\ N^2(X,Y)+  \sqrt{ \f{\ka N(L,Y)}{\b X^{\b\!-\!1}}}\ \nu(X,Y)
 \label{mathX}
\ea
where the noise term has been conveniently renormalized such that 
$\avg{\nu(X,Y)\nu(X',Y')}\! =  \!\delta(Y\!-\!Y')\delta(X\!-\!X')$.

Now, asking for a constant diffusion coefficient 
can easily be achieved by equalizing the  powers of the first two terms of 
\eqref{mathX} namely
\ba 
n\b = 2-2\b \to \b=\f{2}{n+2}
\la{Choice}
\ea
leading  to the equation
\ba
\f{\partial N(X,Y)}{\partial Y}  = \f{A_2}{\b^2}\ \partial_X^2 N(X,Y)\!+\! 
\left(\f{A_2(1-\b)}{\b^2} X^{-1}+\f{A_1}{\b}  
X^{-(1-\b)}\right) \partial_X  N(X,Y)
\nn
+ X^{-2(1-\b)}\left({A_0} N(X,Y) -  N^2(X,Y)\right)   
+  \sqrt{ \f{\ka\ X^{-3(1-\b)}}\b \  N(X,Y)}\ \nu(X,Y) \nonumber \ .
 \label{mathXX}
\ea
Introducing new variables
\[
T = \f {A_2}{\b^2}\ Y, \quad\text{and\ }U(X,T)=\f{N}{A_0}\ ,
\]
Eq. \eqref{mathXX} gets mapped onto the 
``sFKPP- type''
equation 
\ba
\f{\partial U(X,T)}{\partial T} = \partial_X^2 U(X,T)\!+\! 
\left((1\!-\!\b) X^{-1}\!\!\!+\!{\g} 
X^{-(1-\b)}\right) \partial_X  U(X,T)
\!+\! \lambda X^{-2(1-\b)}\left(U\! -\!U^2\right)   
\!+\! \sqrt{\mu\ X^{-3(1-\b)}\ U}\ \nu(X,T) 
 \label{type}
\ea
with
\begin{eqnarray}
&&\g=\b\f{A_1}{A_2}, \quad \lambda=\b^2\f{A_0}{A_2},\quad \mu=\f{\ka}{\b}\ ,\nn
&&\avg{\nu(X,T)\nu(X',T')} = \delta(X\!-\!X')\delta(T\!-\!T')\ .
\la{redef}
\end{eqnarray}
Note that, as well-known, one may 
introduce an extra factor $\sqrt{1-U}$ in the noise term to ensure an absorptive 
boundary also at $U=1$. However, the effects of the 
noise are really important only in the dilute tail of the wave, while this  modification allows to
recover exactly
the 1d sFKPP equation for $\b=1$.

A suggestive interpretation of Eq.\eqref{type} comes from the redefinitions 
\eqref{redef}. While the diffusion constant is by definition rescaled to $1,$ 
the successive coefficient functions acquire the following meaning: 
\bi
\ii
$(1\!-\!\b) X^{-1}\partial X$ is the radial Laplacian term corresponding to a 
medium of noninteger dimension $2\!-\!\b.$
\ii
${\g} 
X^{-(1-\b)}$ is, at $\b=1$, the  ``drift'' term which is now decreasing
at large $X$ for $\b\ne 1$ and thus no more absorbed by a redifinition of 
the ``space'' variable (see \cite{mp}). It may be interpreted  as a kind of 
space-dependent rotation of the  space-time frame.
\ii
$\lambda X^{-2(1-\b)}$ defines a dynamical birth-death coupling strength characterizing a 
medium equipped with a negative gradient; it has thus the important effect of 
a space-dependent ``absorption'' on the wave propagation.
\ii
$\mu\ X^{-3(1-\b)}$ is the new dynamical noise strength,  strongly decreasing 
 when $X$ gets large.
\ei
A qualitative interpretation of the hierarchical behaviour of these coefficients as a function of $X$ suggests  a similar hierarchy of the expected effects on the solutions. Indeed, we will see that it leads to a $\b$-dependent anomalous effective diffusion. Moreover, the fluctuation 
strength (asssociated to the quadratic correlators of $\nu(X,T)$) leads to an 
expected damping of the fluctuations and thus of the stochasticity effects. In the rest of the paper we will thus neglect the stochasticity, postponing its precise study to further work.

It is mathematically interesting to note that Eq.\eqref{mathXX} possesses a well-defined 
limit $n\to\infty,\ i.e.\ \b\to 0,$ giving rise to the equation
\begin{equation}
\label{sfkpp}
\partial_T U(X,T) =  \partial_X^2  U(X,T) + X^{-1}\partial_X  U(X,T)
+\lambda X^{-2} U(1-U) + 
\sqrt{\ka\ X^{-3}U}\ \nu(X,T)\ ,
\end{equation}
which appears to be the {\it radial} sFKPP equation in a 2-dimensional medium 
characterized by a 
{\it negative gradient}  corresponding to the {\it birth-death} factor $\lambda 
X^{-2}.$ The noise 
term is also depressed by the factor $\ka X^{-3}$ which defines the coupling 
strength of the noise correlator.

Hence, in general, for nonzero $\b$ the constant (by construction) 
 diffusive coefficient is larger than the effective birth rate, which is itself 
larger than the effect of the noise correlators. This hierarchy can be 
interpreted as due to a ``radial'' sFKPP  equation in a medium of 
dimensionality between one and two with negative gradient and decreased noise. 
Propagation of waves in such a ``medium''  should and will affect the properties of 
the traveling  solutions which we shall now examine.

\section{3. Scaling solutions}
\la{Solu}

Inspired by the fundamental traveling-wave property of the standard FKPP equation, our strategy 
will be to focus on some region of the forward front and thus  look for a 
scaling variable $\tau(L,Y)$ expressed in the new parameterization as  
$\bx(X,T).$ It will lead to a solution of the form of  a traveling wave 
$U(\bx)$, based on an ansatz valid in a range 
of ``space'' and ``time'' which will be made precise later on. Let us recall that in all the following we shall 
concentrate our study on the deterministic equation without noise.

\subsection{Scaling Solution: First Order Approximation} 
Let us  initially  introduce an  Ansatz 
\ba
U=Ae^{-a\left(X-vTX^{-\al}\right)^b}\ ,
\la{Ans1}
\ea
where the constants $a,b, \al$ and the ``speed'' $v$ have to be determined. Our strategy is
to first  solving Eq.\eqref{type} at leading order when $X\sim \bx$ is 
large w.r.t. $vTX^{-\al}$ and in a second step going beyond this approximation. 
Using the convenient formulae
\ba
\partial_T U = \partial_T \bx \ \partial_{\bx} U \quad \quad \partial_X U = 
\partial_X \bx\ 
  \partial_{\bx} U \quad \quad 
\partial_{X^2}  U=\partial_{X^2}{\bx}\ \partial_{\bx} U + \left(\partial_X 
\bx\right)^2\   \partial_{\bx}^2 U .
\la{useful}
\ea
 we can write, within the same approximation $X\sim \bx$
\ba
\f1U\partial_T U &=&\left(ab\bx^{2b-1}\right)\ v\bx^{-\al}\nn
\f1U\partial_X U &=&\left(-ab\bx^{2b-1}\right)\ \left(1+v\al 
T\bx^{\al-1}\right)\nn
\f1U\partial_{X}^2 U &=&\left(a^2b^2\bx^{4b-2}+ab(1-2b)\bx^{3b-2}\right)\ 
\left(1+v\al T\bx^{\al-1}\right)^2+\nn
&+&\left\{ab\bx^{2b-1}\right\}\ (1-2b)(2-2b)vT\bx^{2b-3}
\ .
\la{need0}
\ea 
In a first step we solve the linear part of Eq.\eqref{type}. As well-known 
\cite{bramson}, the forward front can be studied by restricting first to the 
solution of the linear equation since the nonlinear part has no {\it direct} effect (but 
important {\it indirect} selection effect) on the solution in this range. 
Selecting now leading orders when $X\sim \bx$ is large w.r.t. $vTX^{-\al}$ we find an iterative solution of \eqref{type} by ordering in 
decreasing powers of $X.$ Within the approximation 
$X\sim \bx$ and inserting the relations \eqref{need0} in \eqref{type} one 
realizes that the leading
order should be obtained with terms proportional to $\bx^{2\b-2}.$ 

Collecting 
the leading terms from all contributions one gets
\ba
\f1U \ \partial_T U &:& vab\bx^{2b-\al-1}\nn
\f{1\!-\!\b}{UX}\ \partial_X U &:&-ab(1\!-\!\b)\bx^{2b-2}\nn
\g\bx^{\b-1}\f1U\ \partial_X U &:&-\g ab\bx^{\b+2b-2}\nn
\f1U\ \partial_{X}^2U  &:& a^2b^2\bx^{4b-2}+ab(1-2b)\bx^{3b-2}\nn
\f1U\  \lam U&:& \lam\bx^{2\b-2}
\ .
\la{need00}
\ea 
One thus gets a matching of the dominant contribution in $\bx^{2\b-2}$ by the following 
relations
\ba
\al = 1-\b\ ; \quad b=\f \b 2 \nn
  vab=a^2b^2-\g ab +\lam
\la{order0}
\ea
and thus finds  {\it two} solutions for the parameter $a$, namely
\ba
a_{\pm}=\f1\b\ \left(v+\g \pm \sqrt{(v+\g)^2-4\lam}\right)\ .
\la{Two}
\ea
Hence, as expected from solutions of a second-order equation, the general 
solution
can be (except for the case $a_+=a_-$ see further on) expressed as
\ba
U(X,T) = A_+\ e^{-a_-\left(X-vTX^{\b-1}\right)^{\f {\b} 2}}-A_-\ 
e^{-a_+\left(X-vTX^{\b-1}\right)^{\f{\b} 2}}\ .
\la{Ans2}
\ea
Looking to the solutions \eqref{Two} one sees that one has the condition $v\ge 
2\sqrt\lam$
in order to have real solutions, and thus the dominant speed is minimal when the 
condition
\ba
\la{Cond}
 v_c+\g=2\sqrt \lam = a_c\b\ 
\ea 
is satisfied, where $v_c$ (resp. $a_c$) are  the {\it critical speed} (resp.  
{\it critical slope}). We recognize  among the conditions \eqref{Cond}
the same {\it critical} speed as the one determined by the 
standard method \cite {Brunet,mp} when $n=0,1$ . Note however that the { critical slope} $a_c$
is $\b$-dependent contrary to what is found with the Brunet-Derrida type of Ansatz  \cite {Brunet,mp}.

\subsection{Scaling Solution: Second Order Approximation} 

The critical conditions are realized as a special solution for the ``degenerate case''  $a_+=a_-$  
corresponding to 
the  Ansatz (where we do not add the possible second pure exponential term, as will be justified in the following section) 
\ba
U=Ae^{-a_c\left(X-v_cTX^{\b-1}\right)^{\f\b2}}\ 
\left(X-v_cTX^{\b-1}\right)^{\f\b2}\ ,
\la{Ansaf}
\ea
corresponding to a geometric scaling variable (restoring for a while the $L$
and $n$ dependence), 
\ba
\bx=X-v_cTX^{\b-1}= X-v_cTX^{-\f{n}{n+2}}\equiv L^{\f {n+2}{2}}-v_cTL^{-\f n2}\ .
\la{Vari}
\ea
Taking into account the prefactor in \eqref{Ansaf}, the 
corresponding set of needed expressions (omitting from now on the subscripts  
$v_c\to v;\ a_c\to a$ for simplicity) can be written using for simplicity the 
notation 
 \ba
 \omega = \f X\bx \equiv \left(1-v_cTX^{\b-2}\right)^{-1} \ .
 \la{X}
 \ea
One obtains
\ba
\partial_T U &=& \ U\left(a\b/2\bx^{\b-1}-\b/2\bx^{\b/2-1}\right)\ 
v\bx^{\b-1}\omega^{\b-1}\nn
\partial_X U &=& \ U\left(-a\b/2\bx^{\b-1}+\b/2\bx^{\b/2-1}\right)\ 
\left(1+v(1-\b) T\bx^{\b-2}\omega^{\b-2}\right)\nn
\partial_{X}^2 U &=& \ U\left(a\b/2\bx^{\b-1}-\b/2\bx^{\b/2-1}\right)^2\ 
\left(1+v(1-\b) T\bx^{\b-2}\omega^{\b-2}\right)^2+\nn
&+& \ U\left(a\b/2(1-\b)\bx^{\b-2}-\b/2(1-\b/2)\bx^{\b/2-2}\right)\ 
\left(1+v(1-\b) T\bx^{\b-2}\omega^{\b-2}\right)^2+\nn
&+& \ U\left\{a\b/2\bx^{\b-1}-\b/2\bx^{\b/2-1}\right\}\ 
(1-\b)(2-\b)vT\bx^{\b-3}\omega^{\b-3}
\ .
\la{needf}
\ea 
Now, we find an iterative solution of \eqref{type} by ordering in decreasing 
powers of $X.$ Assuming before further discussion to consider the kinematical 
region where
\ba
\omega = 1+{\cal O}\left(T\bx^{\b-2}\right)\sim 1\ ,
\la{assum}
\ea 
and inserting the relations \eqref{needf} in \eqref{type} order by order, and   
considering  the leading terms from all contributions one gets
\ba
\f1U \ \partial_T U &:& va\b/2\bx^{2\b-2}-v\b/2\bx^{3\b/2-2}+\cdots\nn
\f{1\!-\!\b}{UX}\ \partial_X U 
&:&-a\b/2(1\!-\!\b)\bx^{\b-2}+\b/2(1\!-\!\b)\bx^{\b/2-2}+\cdots\nn
\g\bx^{\b-1}\f1U\ \partial_X U &:&-\g 
\left(a\b/2\bx^{2\b-2}-\b/2\bx^{3\b/2-2}\right)+\cdots\nn
\f1U\ \partial_{X}^2U  &:& 
a^2\b^2/4\bx^{2\b-2}-a\b^2/2\bx^{3\b/2-1}+\left(a\b/2(1-\b)-\b^2/4\right)\bx^{\b
-2}-\b/2(1-\b/2)\bx^{\b/2-2}+\cdots\nn
\f1U\  \lam U&:& \lam\bx^{2\b-2}
\ .
\la{needg}
\ea 
Collecting the leading terms from all contributions to Eq.\eqref{mathXX} one gets 
from the four first orders
\ba
\bx^{2\b-2}&:&\ a^2\b^2/4-va\b/2-\g a\b/2 +\lam=0\nn
\bx^{\f 32\b-2}&:&\ v\b/2+\g \b/2-a \b^2/2 =0 \nn
\bx^{\b-2}&:&\ -a\b/2(1\!-\!\b)+a\b/2(1\!-\!\b)-\b^2/4 =-\b^2/4 \nn
\bx^{\f 12\b-2}&:&\ \b/2(1\!-\!\b)-\b/2(1\!-\!\b/2) =-\b^2/4 \ 
 .
\la{orderg}
\ea
 Both first relations are satisfied by the {\it critical condition}  \eqref{Cond}
$v_c+\g=2\sqrt \lam = a_c\b$. Hence the consideration of the 
special solution \eqref{Ansaf} allowed for one  subleading order veryfying 
the equation. The  nonzero remaining 
orders are $\bx^{\b-2}, \bx^{\f 12\b-2}$. As we shall see further on, the subsubleading 
contributions, together with those coming beyond the approximation \eqref{assum}
 will be dealt with the linearly $T$-dependent terms. Interestingly 
enough, these subsubleading powers will be related, in the known cases at least 
\cite{bramson,mp}, to the subleading terms of the universal expansion of the 
scaling variable.


\section{4. Extending Geometric Scaling}
\la{Enhan}
In order to explore the kinematic range of validity of the scaling variable $\bx 
(X,T)$ and look for corrections allowing to take into account
sub-subleading corrections induced by the set of equations  \eqref{orderg}, let 
us recall the previously known results on geometric scaling coming for the 
studies \cite{mp} on nonrunning ($n=0$) and running  ($n=1$) BK equations and make the appropriate  generalization of these known results for generic $n.$ 

One is led to distinguish  3 regions \cite{bramson} (we will see that we can add a 4th one) namely, in their 
order towards  the moving tail of the wave, the {\it wave interior} region, where geometric  
scaling  has been identified, the  {\it leading edge}, where diffusive 
corrections appear, and the  {\it very forward} region where the nonuniversal 
initial conditions prevail. Extrapolating for all $n$ the known results for $n=0\ {\rm and}\ 
1,$ the wave interior region can be  characterized by the kinematic relation $L\sim 
(v_cT)^{\f1{n+1}}$ and the leading edge region by $L\sim 
(v_cT)^{\f1{n+1}}+\rho(v_cT)^{\f1{n+2}}$ where $\rho \sim cst.$ 

Focussing on  the  {\it leading edge} and extrapolating for all $n$
 the Ansatz proposed in refs. \cite {Brunet,mp} for the standard running and 
nonrunning FKPP cases, the generic solution reads
\ba
U_n(X,T) \propto  T^{\f 1{n+2}}\ G_{n}\left(
\f{L-(v_cT)^{\f1{n+1}}+c_n(T)}{T^{\f 1{n+2}}}\right)\ 
e^{\g\{L-(v_cT)^{\f1{n+1}}+c_n(T)\}}\ ,
\la{Leading}
\ea
where $n=(0,1)$ is the previously defined index for nonrunning and running 
cases, $\tau_n(L,T)\equiv L-(v_cT)^{\f1{n+1}}+c_n(T)$ is the  original scaling variable 
extrapolated from the values obtained in \cite{mp}, $c_{n}(T)$ the corresponding 
subleading universal contributions to the scaling variable and $\g$ a universal 
critical slope value characterizing the scaling in the {\it wave interior} when 
$ z \equiv 
{\tau_n(L,T)}/{T^{\f 1{n+2}}}\to 0.$ Note the major constraint in this limit 
$G_{n}(z)\sim z$ which expresses the ``absorptive'' property \cite{Abs} due to 
the nonlinear terms. This is the very same property which is realized in our Ansatz \eqref{Ansaf}, Motivating the absence of a pure exponential term $a\ priori$ allowed for a solution of a second-order linear equation.

The function $G_{n}$ has been determined for $n=0,1$ (see 
$e.g.$ \cite{mp}), while their precise form in the general case, not relevant 
for our needs, should not be difficult to be obtained in a similar way.

Within the typical leading edge region where $L\sim 
(v_cT)^{\f1{n+1}}-c_n(T)\tau+\rho_n T^{\f 1{n+2}},$ where $\rho_n$ is a smooth  function around some constant value. We see that the 
expression \eqref{Leading} takes the  $scaling$ form
 \ba
U_n(X,T) \propto T^{\f 1{n+2}}\ G_{n}\left(\rho_n\right)\ e^{\g\rho_n T^{\f  
1{n+2}}}\sim \f1{\rho_n}\tau_n(L,T)\ G_{n}\left(\rho_n\right)\ e^{\g\tau_n(L,T)}\ .
\la{Leading1}
\ea
Hence in the leading edge region, we can formulate the standard solution   with essentially the same original scaling variable. 

Strikingly enough, a matching with our obtained solution \eqref{Ansaf} is 
obtained  in an appropriate kinematic region where
\ba
X \sim 
Kv_cT\ ; \quad\quad L \sim (Kv_cT)^{\f{n+2}2},
\la{Id1}
\ea
 where the constant $K>1$  in the nonrunning case.
This kinematic region is always near and forward to the identified {\it leading 
edge} regions  characterized (see \cite{mp} for $n={0,1}$) by 
$L\sim v_c T^{\f1{n+1}}+{\cal O}(T^{\f 1{n+2}}).$ For instance, 
$L\sim (v_c+K)T$ for the nonrunning case and $L\sim v_c T^{2/3}$
for the running case. Let us call for convenience this region the $extended\ 
scaling$ region \cite{footnote}.  Note also that  one identifies
\ba
a_c \equiv {(n+2)\sqrt{\lambda}}\sim 
\g\rho_n \ .
\la{Id}
\ea

Hence requiring a matching between the $leading\ edge$ and $extended\ scaling$ 
regions, and using the scaling variable $\bx\equiv X-v_cTX^{\f n{n+2}}$ leads to 
an extended scaling domain. 
Interestingly enough the scaling variable gets modified, since
\ba
\bx \equiv X-v_cTX^{-\f1{n+2}}=L^{\f{n+2}2}-v_cTL^{-\f n2}\ .
\la{Modif}
\ea
As an application to the physical cases, in the nonrunning case, one realizes that the 
scaling variable is the original one, while the overall scaling region 
has been enhanced thanks to the modification of the scaling function. In the running coupling case, one finds $X-v_cTX^{-\f1{3}}=L^{\f{3}2}-v_cTL^{-\f12},$  which shows the compatibility with the standard scaling variable $\tau=L-(v_cT)^{1/2},$  since  $L^2 \sim v_cT$ near the wave front.

We thus obtain a new scaling solution extended in the forward region matching at its lower $X$ boundary with the {\it 
leading edge}. We thus infer that, 
for all value of $\b=\f 2{2+n}$ the  variable  $X-v_cTX^{1-\b}
\equiv L^{\f n2+1}-v_cTL^{-\f n2}$
is a promising candidate to ensure an extended geometric scaling.

 In order to check the validity of the scaling in this domain, one has to ensure that the 
approximation \eqref{assum} is valid for $\bx\propto T.$ By inspection of 
the system of equations \eqref{needf}, one realizes that the correction is of 
order $\bx^{3\b-4}<< \bx^{2\b-2}\bx^{\f32\b-3}$ at large $\bx,$ ensuring that 
the two first relations of \eqref{orderg} are still valid, preserving the 
enhanced scaling property. Hence beyond the {\it wave interior} and the {\it leading edge} and before the {\it very forward} regions classically analyzed $e.g.$ in Ref.\cite{bramson}, one finds a new {\it extended scaling} region
for generalized FKPP equations and their traveling wave solutions.

Concerning the subsubdominant orders in \eqref{orderg}, they may receive 
contributions from those  $3\b-3$ terms coming from \eqref{assum}  (except for the nonrunning case $\b=1$) and those in 
$\b-2$ and $\f\b2-2.$ A simple way to take these terms into account is to add 
subleading terms in 
the $T$-dependence of the scaling variable taken into account by the  function 
$c(T)$ which appears in the {\it leading edge} formula \eqref{Leading}. For 
instance, considering
\ba
\bx=X-v_cT+h\log T+kT^{-\f 12}+\cdots\ for\ \b= 1 \ ; \quad\quad\bx=X-v_cT+hT^{1-\b}+kT^{-\f {3\b}2}+\cdots\ for\  \ 
\b< 1\ ,
\la{Uni}
\ea
one generates the  order $\b-2$ and $\f\b2-2$.  Other inputs will easily ensure the cancellation of the subdominant terms of order $3\b-3$ present in the $\b \ne 1$ cases. Note the interesting fact that  the expressions \eqref{Uni} are exactly  matching   the orders of the well-known subdominant universal terms \cite{bramson,mp}. In the case  $\b \ne 1$, new terms appear which deserve a specific study. We postpone this  for further  work.


\section{5. Summary and Outlook}
\la{Sum}
Let us summarize our main results:
\bi
\ii
{\it Mapping to (s)FKPP universality classes.} We introduce a one-parameter 
indexed by $n$ family of nonlinear equations including and interpolating  the 
Balitsky-Kovchegov equations with nonrunning and running QCD couplings as 
$\{\log Q^2\}^{-n}.$ In the diffusive approximation, this family is exactly 
mapped to a (s)FKPP equation describing the propagation of radial traveling 
waves in an absorptive medium of dimensionality ${2(n+1)}/{(n+2)}$ and coefficient 
functions depending on the ``space'' variable $X=\{\log Q^2\}^{\f2{n+2}}$. The 
different coefficients verify (except for $n=0$) at large $X$ the hierarchy
\ba
X^0\ (Diffusion) >> X^{-\f n{n+2}}\ (``Drift'') >> X^{-\f{2n}{n+2}}\ 
(Birth/Death\ rate) >> X^{-\f{3n}{(n+2)}}\ (Noise\ strength)
\la{Hier}
\ea 
\ii
{\it Geometric Scaling Solutions.} Using an expansion at large $X$, we find a new
geometric-scaling   
domain forward to the canonical traveling wave front such that $L\sim 
(Kv_cT)^{\f2{n+2}},$ where $T\propto Y$ the rapidity plays the role of 
``time'',  $v_c$ is the minimal speed (the constant $K$ is  arbitrary
except that $K>1$ when
 $n=0$)
with the scaling solution 
\begin{equation}
U\propto\ e^{-(n+2)\sqrt {\lambda}\left(L^{\f{n+2}2}-v_cTL^{-\f n{2}}\right)^{\f 1{n+2}}}\  \left(L^{\f{n+2}2}-v_cTL^{-\f n{2}}\right)^{\f 1{n+2}}\ ,
\la{Ansafinal}
\end{equation}
where $\lambda$ is a constant depending on the BK equation parameters, see 
\eqref{redef}.
\ii
{\it Extended Geometric  Scaling.}
Matching with the universal solutions  in the near-by less forward {\it leading 
edge} region, the scaling variable $L^{\f{n+2}2}-
v_cTL^{-\f n{2}}$ is shown to allow for 
an extension of the  geometric-scaling region in a new kinematic domain. 

Specifying for the nonrunning case ($n=0$), 
one has the same scaling variable as previously, but a new scaling form allowing 
to extend the geometric scaling region from the wave interior  $L\sim v_cT,$ 
including  the leading edge region with $L\sim (v_cT)+ \rho_0 T^\f 1{2}$ up to 
an extended forward region where $L\sim Kv_cT$ with $K>1.$ 

For the running and more general $n\ne 0$ cases, one finds starting from the 
{\it wave interior} region $L\sim v_cT^{\f 1{n+1}},$ scaling in  the {\it leading edge }
region $L\sim (v_cT)^{\f 1{n+1}}+ \rho_1 T^\f 1{n+2}$ up to an {\it extended scaling} 
region where $L\propto (v_cT)^{\f2{n+2}}.$
\ei 
The output of these properties is interesting in particular for the 
running coupling case ($n=1$), which is   physically meaningful for the BK 
equation in QCD. For instance, the parameter  $a_c=(n+2)\sqrt \lambda$ related to an 
effective slope of the scaling curve  is increasing by a factor $\f32$ from 
nonrunning to the running case, while the leading edge form leads to a constant 
slope. It would be worth comparing this prediction with a simulation of the BK 
equation \cite{Albacete:2007yr} where a similar effect has been noticed.

On the phenomenological ground, which is not our subject in the present paper, there are quite a few issues which deserve to be studied. Last but not least, it would be worthwhile to confront the new extended scaling properties, in particular the new scaling variable $\bx=X-v_cTX^{-\f13}\equiv 
L^{\f32}-(v_cT)L^{-\f12}$, with data using $e.g.$ the method of 
Refs.\cite{Gelis:2006bs}. Also, geometric scaling in other reactions, such as hard diffractive events \cite{Diff}, would be welcome.

On the theoretical ground, some studies remain to be done in the new framework. For instance, the effect of a cut-off \cite{Brunet} could be analyzed using the general solution \eqref{Ans2} with probably complex conjugate functions and match with the cut-off region. Also, the investigation of subleading universal terms is possible, thanks to appropriate modification of the time dependence of the scaling variable. It could be worthwhile to merge the new scaling approach with the studies on the parametric form of the traveling waves \cite{Para}.

One would also like to go beyond the diffusive aproximation. At the present stage of our theoretical knowledge, the BK equation has been corrected for the full nest-leading log contributions \cite{NLL}. Indeed, the matching with the standard universal formulation ensures that one should not expect dramatic effects, except for the neat difference between the fixed and running coupling cases \cite{Univ}, but this dedicated study deserves to be done in the future. In any case, it would be interesting to see whether and how it works. 
                                                                                                                                                                        \section*{Aknowledgements}
The autor would like to thank Guillaume Beuf, Andrzej Bialas and Cyrille Marquet for useful discussions and (for C.M.) a careful critical reading of the manuscript. During the completion of this work, he  was partly supported by  the ``Partenariat Hubert Curien - Polonium 2009'' by the Ministry of Foreign and European Affairs, France, and  thanks the Institute of physics of the Jagiellonian university in Cracow for hospitality.

\end{document}